\magnification=1200
\font\stepfour=cmr10 scaled\magstep3

\font\ninerm=cmr9
\hfuzz=12pt
\baselineskip=15pt
\def\makefootline{\baselineskip=36pt\line{\the\footline}}
\nopagenumbers 

%
\catcode`\*=11  

\def\sl*sh#1#2#3{\ooalign{\setbox0=\hbox{$#2\not$}
                          $\hfil#2\mkern-24mu\mkern#1mu
                           \raise.15\ht0\box0\hfil$\cr
                          $#2#3$}}
\catcode`\*=12 

\rightline{UCLA/96/TEP/13}

\bigskip\bigskip

\centerline{\bf\stepfour Worldline Supersymmetry and Dimensional Reduction }

\bigskip\bigskip\bigskip\bigskip

\centerline{Darius G. Gagn\'e}
\medskip
\centerline{\it Department of Physics and Astronomy}
\medskip
\centerline{\it University of California, Los Angeles}
\medskip
\centerline{\it Los Angeles, CA 90024, USA}
\medskip
\centerline{\it gagne@physics.ucla.edu}

\bigskip\bigskip\bigskip\bigskip\bigskip\bigskip

\centerline{\bf Abstract}

\medskip

{\ninerm{\noindent  For 
any worldline reformulation of a quantum field theory for 
Dirac fermions, this paper shows that 
worldline supersymmetry may generally be enforced by the
vanishing of the commutator of the Dirac operator 
with the worldline Hamiltonian.  
The action of supersymmetry on the worldline Lagrangian may not,
however, be written in terms of the variations on the fields in the
usual way, except when the spinning particle couples just to a
one-form.  By reduction 
{}from six to four dimensions of the worldline reformulation for a spinning
particle coupled to a three-form, corrections to the superworldline
Lagrangian are presented which are needed in order to reproduce
correct field theory results from worldline perturbation theory 
in an unambiguous way. }}

\vfill
\eject

\footline={\hss\tenrm\folio\hss}

\noindent
{\bf 1. Introduction}

\bigskip

As the point particle limit of superstring theory, the worldline formalism
for spinning particles is expected to have a $ D = 1 $, $ N = 1 $ 
supersymmetry or worldline supersymmetry.  This has been shown to be true for
a spinning particle coupled to an Abelian [1] and a non-Abelian 
[2] 
vector gauge field.  For a review of supersymmetric quantum mechanics,
see ref.\ [3].  The worldline Lagrangians for a spinning particle
coupled to a scalar and  pseudoscalar [4] as well as to a 
vector and axial vector [5] (all four fields Abelian) 
have been constructed in a 
manifestly supersymmetric way as the description of a spinning particle
on a superworldline [1], 
although the actual supersymmetry transformations have not been discussed.

Recently 
[6], the worldline reformulation for a multiplet of Dirac fermions 
coupled to the most general set of non-Abelian 
background scalar, pseudoscalar,
vector, axial vector and antisymmetric tensor fields has been derived from
field theory.
For the Abelian scalar, pseudoscalar and vector, the 
derived Lagrangian agrees with the result of [5], while for the Abelian
axial vector there is agreement provided the auxiliary 
fields coming from the superworldline Lagrangian are treated in a somewhat 
ambiguous way.
However, when including an antisymmetric tensor or a non-Abelian axial
vector, the derived Lagrangian 
curiously contains terms which are not expected to appear in  
a manifest superworldline construction but are needed in order 
to produce correct field theory results.   In other words, the full
worldline Lagrangian derived from field theory
is apparently not supersymmetric.  
Why should some worldline Lagrangians be supersymmetric and others turn out 
not to be?

By defining supersymmetry transformations 
on the worldline in a more general
way than the superworldline construction, this paper shows in the 
next section that the spinning
particle Lagrangian is actually {\it always} supersymmetric for any background 
coupling.  Specifically, identification of the Dirac operator 
with the supercharge is made and
the action of supersymmetry on the worldline Hamiltonian
is given simply by its commutator
evaluated at $\hbar = 0$ with the supercharge.
It is shown that the action of supersymmetry on the Lagrangian
can not be given by the usual functional variation, except when the spinning
particle couples just to a one-form.

The last section investigates why the superworldline Lagrangian becomes
problematic for the coupling to an axial vector or to an 
antisymmetric tensor.  It is shown
that the superworldline construction for
a spinning particle coupled in any dimension to higher
p-forms generally does not produce a number of terms present
in the corresponding Lagrangian derived from field theory,
except for the special case ${\rm p} = 1$.  
{}From the dimensional reduction of the worldline 
reformulation for a spinning particle
coupled to a three-form in six dimensions,  the superworldline
Lagrangian for the coupling to an Abelian axial vector and an antisymmetric 
tensor, expressed in terms of the auxiliary fields
introduced in [4 and 5], 
is recovered {\it plus} the correction terms which render the whole 
expression both correct and 
invariant under the {\it new} supersymmetry transformations.  
Finally, these insights make it possible to give simple unambiguous rules for
all worldline perturbation theory, including expressions involving 
auxiliary fields. 

\bigskip

\noindent
{\bf 2. Worldline Supersymmetry}

\bigskip

Consider the worldline reformulation of a quantum field theory for a Dirac
fermion with general couplings in $D$-dimensions, analytically continued to
Euclidean space.  The Dirac operator,  
$\hat{\Sigma}$, should appropriately
be taken without loss of generality
to be hermitian and to be of fermionic grading
\footnote{${}^\dagger$}{Such a Dirac operator is required
in a worldline reformulation and may always be obtained from
a general one by doubling the fermionic degrees of freedom 
[6 and 7].}.
{}From refs. [6 and 7], $\hat{\Sigma}$ may generally be cast as
$$
\hat{\Sigma} = \Gamma_A \hat{p}_A - \sum_{r=0} 
i^r \Gamma_{A_1 \dots A_{2r+1} } \hat{K}^{(2r+1)}_{A_1\dots A_{2r+1}} ~ ,
\eqno (2.1)
$$
where $\hat{K}^{2r+1} = K^{2r+1}(\hat{x})$ is an hermitian 
Abelian $(2r+1)$-form,
$\Gamma_A$, $A=1,\dots ,D$ 
are the hermitian generators of the Euclidean Clifford algebra in
$D$ dimensions, $ \Gamma_{A_1 \dots A_{2r+1} }$ is the totally antisymmetric
product of $\Gamma_{A_1},\dots , \Gamma_{A_{2r+1}}$ and $\hat{p}_A$ is the 
hermitian momentum operator.  The spinning particle worldline Lagrangian
is then
$$
L = L_K + H ~ ,
\eqno (2.2a)
$$
where the kinetic part of the Lagrangian, $L_K$, is given by
$$
L_K = -i\dot{x}_A p_A + {1\over 4}\psi_A \dot{\psi}_A 
\eqno (2.2b)
$$
and the worldline Hamiltonian, $H$, is obtained from the Hamiltonian operator 
$$
\hat{H} = \hat{\Sigma}^2 
\eqno (2.2c)
$$
in the limit $\hbar\to 0$.
``Observable'' classical quantities are obtained from the quantum
mechanical operators in the $\hbar \to 0$ limit by the rules
$$
\eqalignno{
\Gamma_{A_1 \dots A_{2r+1} } 
&\to \psi_{A_1} \cdots \psi_{A_{2r+1}} & (2.3a) \cr
\{ \hat{p},F(\hat{x})\} 
&\to 2pF(x) ~ .                        & (2.3b) \cr}
$$
Now, since $\hat{\Sigma}$ is hermitian and fermionic, it may be identified 
as the supercharge generating the
$D=1$ supersymmetry algebra of eqn.$(2.2c)$.  Use will now be made of 
the fact that the Hamiltonian
commutes with the supercharge,
$$
[\hat{H},\hat{\Sigma}] = 0 ~ ,
\eqno (2.4)
$$  
in order to define worldline supersymmetry in a general way.
First, supersymmetry transformations on the fields are defined as
$$
\eqalignno{
\delta_\alpha x_A 
&= [\hat{x}_A,\alpha\hat{\Sigma} ]_{(\hbar\to 0)} 
 = i\alpha\psi_A						& (2.5a) \cr
\delta_\alpha p_A
&= [\hat{p}_A,\alpha\hat{\Sigma} ]_{(\hbar\to 0)} 	
 = i\alpha\sum_{r=0} i^r \psi_{A_1} \cdots \psi_{A_{2r+1}}
   \partial_A K^{(2r+1)}_{A_1\dots A_{2r+1}}			& (2.5b) \cr
\delta_\alpha \psi_A 
&= [\Gamma_A,\alpha\hat {\Sigma}  ]_{(\hbar\to 0)} 
 = -2\alpha p_A + 2\alpha \sum_{r=0} 
   i^r \psi_{A_2} \cdots \psi_{A_{2r+1}} 
   K^{(2r+1)}_{AA_2 \dots A_{2r+1}} ~ ,  			& (2.5c) \cr} 
$$
where $\alpha$ is the Grassmann parameter for the supersymmetry 
transformations.  
These transformations may equivalently be defined in terms of the Poisson
bracket \footnote{${}^\dagger$}{This bracket may more correctly be 
referred to as a Dirac bracket [8].}
$$
\delta_\alpha f
= [f,\alpha\Sigma ]_{\rm{P.B.}} ~~~ , ~~~ f = x,p ~ \rm{or} ~ \psi ~ ,
\eqno (2.6a)
$$
defined as 
$$
[A,B]_{\rm P.B.} 
= i{\partial A\over\partial x_A}{\partial B\over \partial p_A}
- i{\partial A\over\partial p_A}{\partial B\over \partial x_A}
-2\Bigl( \delta_{0B}(-)^{A+B} + \delta_{1B} \Bigr)
{\partial A\over \partial \psi_A}{\partial B\over \partial \psi_A} ~ .
\eqno (2.6b)
$$
These brackets may be shown to define a graded Lie algebra [8].  
We have also
introduced the observable supercharge
$$
\Sigma = \hat{\Sigma}_{\hbar\to 0} =  \psi_A p_A - \sum_{r=0} 
i^r \psi_{A_1} \cdots \psi_{A_{2r+1}} K^{(2r+1)}_{A_1\dots A_{2r+1}} ~ ,
\eqno (2.7)
$$
where use has been made of the rules in $(2.3)$.

Now, under the variations of $(2.5 ~ \rm{or} ~ 6)$, 
the kinetic part of the Lagrangian
is generally invariant (up to a total derivative in the propertime) with 
its variation defined as
$$
\delta_\alpha L_K 
\equiv L_K 
(x + \delta_\alpha x , p + \delta_\alpha p , \psi + \delta_\alpha \psi )
- L_K (x,p,\psi ) 
= \partial_\tau 
({1\over 4}\psi_A\delta_\alpha\psi_A 
- ip_A\delta_\alpha x_A - \alpha\Sigma ) ~ .
\eqno (2.8)
$$  
The action of supersymmetry on 
the Hamiltonian is defined to be
$$
\delta_\alpha H \equiv [\hat{H},\alpha\hat{\Sigma}]_{(\hbar\to 0)} ~ ,
\eqno (2.9)
$$
which must vanish by (2.4).  Thus quite generally, the worldline 
Lagrangian, $L$, given by $(2.2a)$, is invariant up to a total derivative 
under the supersymmetry transformations (2.8, and 9).  However, it will
now be demonstrated that the variation of the Lagrangian, $L$, is not
given by the usual functional variation, i.e.,
$$
\delta_\alpha L
\not= 
L (x + \delta_\alpha x , p + \delta_\alpha p , \psi + \delta_\alpha \psi )
- L (x,p,\psi ) ~ ,
\eqno (2.10)
$$
except when the fermion couples just to a one-form.

In order to demonstrate (2.10), it suffices to show that the variation
of the Hamiltonian, $H$, is not given by the usual functional variation
in terms of the variations on the fields, 
$\delta\alpha x$, $\delta_\alpha p$ and $\delta_\alpha \psi$.
So, to see what the transformation (2.9) on the Hamiltonian
looks like in terms of the fields,
it is instructive to first consider the 
coupling
just to the one-form ($r=0$) in $(2.1)$, which gives
$$
H = (p - K)^2 + i \psi_A \psi_B \partial_A K_B ~ .
\eqno (2.11)
$$
It is easy now to check by  $(2.5 ~ \rm{or} ~ 6)$ that 
$$
\delta_\alpha H 
= 
H (x + \delta_\alpha x , p + \delta_\alpha p , \psi + \delta_\alpha \psi )
- H (x,p,\psi )  
= [H,\alpha\Sigma ]_{\rm P.B.} ~ \Bigl( = 0 \Bigr) ~ .
\eqno (2.12)
$$
In particular, the
commutator $[\hat{H},\alpha\hat{\Sigma}]_{(\hbar\to 0)}$ and the
Poisson bracket  
$[H,\alpha\Sigma ]_{\rm P.B.}$ are
identical term by term (and of course these terms add to zero).
For example, 
$$
[i\Gamma_{AB}\partial_A\hat{K}_B , \alpha\Gamma_C\hat{p}_C]_{(\hbar\to 0)}
=
[i\psi_A\psi_B\partial_AK_B ,\alpha\psi_Cp_C ]_{\rm P.B.}
=
2i\alpha(p_A\psi_B\partial_BK_A - p_A\psi_B\partial_AK_B ) ~ .
\eqno (2.13)
$$
The one-form coupling may be used 
for example to obtain
the usual gauge, scalar and 
pseudoscalar couplings in four dimensions by reduction from dimension 
$D\geq 6$ [4].  

Next, consider the coupling
just to the three-form ($r=1$) in $(2.1)$, which gives
$$
\eqalign{
H 
&= p^2 - 6i\psi_A\psi_B p_C K_{ABC} 
- \psi_A\psi_B\psi_C\psi_D\partial_A K_{BCD}
\cr
&+ 6 K_{ABC}K_{ABC} - 9 \psi_A\psi_B\psi_C\psi_D K_{EAB}K_{CDE} ~ . 
\cr}
\eqno (2.14)
$$
Now notice that 
$$
H (x + \delta_\alpha x , p + \delta_\alpha p , \psi + \delta_\alpha \psi )
- H (x,p,\psi ) 
=
[H,\alpha\Sigma ]_{\rm P.B.} 
= 12i\alpha\psi_A\partial_A K_{BCD} K_{BCD} \not= 0 ~ .
\eqno (2.15)
$$
Thus, curiously  
$$
[\hat{H},\alpha\hat{\Sigma}]_{(\hbar\to 0)} 
\not= [H,\alpha\Sigma ]_{\rm P.B.}
\eqno (2.16)
$$
in general.  Thus the supersymmetry transformation (2.9) for the Hamiltonian
is not given simply by the usual functional variation (i.e. Poisson bracket)
as in (2.12) except for the special case when a fermion couples just to  
a one-form.
This surprise can be understood in the following way.  The commutator and
the Poisson bracket both agree on the term that generates (2.15):
$$
[\hat{K}_{ABC}\hat{K}_{ABC},\alpha\Gamma_D\hat{p}_D]_{(\hbar\to 0)}
=
[{K}_{ABC}{K}_{ABC},\alpha\psi_Dp_D]_{\rm P.B.}
=
12i\alpha\psi_A\partial_A K_{BCD} K_{BCD} ~ .
\eqno (2.17)
$$
For the commutator, the cancellation of (2.15 or 17) comes from a piece of
the term 
$$
\eqalign{
[-\Gamma_{ABCD}\partial_A\hat{K}_{BCD},
&-i\alpha\Gamma_{EFG}\hat{K}_{EFG}]_{(\hbar\to 0)} 		\cr
&\sim
-12i\alpha\psi_A\partial_A K_{BCD} K_{BCD} 
+36i\alpha\psi_A\partial_B K_{ACD} K_{BCD}
+\cdots ~ .							\cr}
\eqno (2.18)
$$
The commutator and Poisson bracket 
agree perfectly on the terms denoted as $\cdots$.  These 
terms arise due to a single contraction of $\Gamma$-matrices under the
commutator, which is equivalent to the single derivative by worldline fermions
on each argument under the Poisson bracket (2.6).  However, the first two
terms in (2.18) arise due to three contractions of $\Gamma$-matrices under
the commutator, which at the observable level is equivalent to three
derivatives by worldline fermions on each argument.  Such higher derivative
terms are obviously not accommodated by the Poisson bracket (2.6).
Moreover, the second term in (2.18) is cancelled at the 
commutator level by a piece of the term 
$$
[-3i\Gamma_{AB}\{ \hat{p}_C , \hat{K}_{ABC} \} ,
-i\Gamma_{DEF}\hat{K}_{DEF}]_{(\hbar\to 0)}
\sim
-36i\alpha\psi_A\partial_B K_{ACD} K_{BCD}
+ \cdots ~ .
\eqno (2.19)
$$
This leading term can again only be obtained at the observable level
by higher derivatives lacking in the definition of the Poisson bracket in
(2.6).  

It is possible,
however, 
to define a generalized Poisson bracket (G.P.B.) as a formula for 
$\delta_\alpha H$ in terms of observables by analyzing how the higher
contractions
between $\Gamma$-matrices go in the commutator of (2.9).  For the case of 
a three-form, the higher contractions are given 
specifically by (2.18) and (2.19) and
so the (vanishing) variation of the Hamiltonian may be cast as
$$
\eqalign{
\delta_\alpha H 
&= [H,\alpha\Sigma ]_{\rm P.B.} 
+ 
{i\over 2}
{\partial^3 H             \over \partial\psi_A\partial\psi_B\partial p_C}
{\partial^3(\alpha\Sigma )\over \partial\psi_A\partial\psi_B\partial q_C}
+
{1\over 3}
{\partial^3 H             \over \partial\psi_A\partial\psi_B\partial\psi_C}
{\partial^3(\alpha\Sigma )\over \partial\psi_A\partial\psi_B\partial\psi_C}
\cr
&\equiv  [H,\alpha\Sigma ]_{\rm G.P.B.} ~ .
\cr} 
\eqno (2.20)
$$ 
This generalized Poisson bracket generates the same equations of 
motion and supersymmetry transformations for the fields 
$x,p ~ {\rm and } ~ \psi$ as the Poisson bracket (2.6).  However,
the generalized Poisson bracket does not obey the super Jacobi identity.

This situation becomes increasingly more severe as the fermion couples to
higher forms.  Variation of the Hamiltonian by the 
Poisson bracket leaves numerous terms 
uncancelled.  These terms can only be cancelled 
using the correct variation formula (2.9) or equivalently 
variation by a generalized Poisson bracket with a suitably high
number of derivatives.

\bigskip

{\bf 3. Superworldline and Dimensional Reduction}

\bigskip

The fact that $\delta_\alpha H \not= [H, \alpha\Sigma ]_{P.B.}$ except when
the fermion couples to a one-form ($r=0$) will now be shown to mean 
equivalently that the worldline Lagrangian $L$ of $(2.2a)$ derived from 
field theory differs from the Lagrangian obtained from the 
superworldline construction, except when $r=0$.
Firstly, using the equations of motion for the momentum, the supersymmetry 
transformations (2.5) on the fields become
$$
\eqalignno{
\delta_\alpha x      &=  i\alpha \psi_A 	&  (3.1a)	\cr
\delta_\alpha \psi_A &= -i\alpha \dot{x}_A ~ .	&  (3.1b)	\cr}
$$
The same variations on the fields can be obtained by considering a
superworldline $( \tau , \theta )$, where $\theta$ is a Grassmann 
number.  By defining the superfield $X$, supercharge $Q$ and 
superderivative $D$ as
$$
X_A = x_A + \theta\psi_A ~~~ , ~~~ 
Q   = {1 \over i}(\partial_\theta + \theta\partial\tau) ~~~ {\rm and} ~~~
D   = {1 \over i}(\partial_\theta - \theta\partial\tau) ~ ,
\eqno (3.2)
$$
supersymmetry variations (3.1) are reproduced by
$$
\delta_\alpha x_A = \int d\theta \theta ~ [X_A , \alpha Q]
~~~ {\rm and} ~~~
\delta_\alpha \psi_A = \int d\theta  ~ [X_A , \alpha Q] ~ .
\eqno (3.3)
$$

Now, the Lagrangian $(2.2a)$ 
for the coupling to just a 
one-form ($r=0$) with the momentum integrated
out becomes 
$$
L = { \dot{x}^2 \over 4 } + {1 \over 4}\psi_A\dot{\psi}_A  - i\dot{x}_A K_A
  + i \psi_A \psi_ B \partial_A K_B ~ .
\eqno (3.4)
$$
This Lagrangian can be formulated as a superworldline Lagrangian,$L_s$,
$$
L_s 
\equiv 
\int d\theta \Bigl( {1 \over 4i}DX_AD^2X_A - DX_AK_A \Bigr) 
= L ~ .
\eqno (3.5)
$$
This Lagrangian is invariant under supersymmetry transformations (3.1) with 
simply a functional variation: 
$$
\delta_\alpha L
\equiv
L(x + \delta_\alpha x,\psi + \delta_\alpha \psi) - L(x,\psi)
=
\partial_\tau \Bigl[ \alpha\psi_A\bigl(
K_A - {i \over4}\dot{x}_A \bigr) \Bigr] ~ .
\eqno (3.6)  
$$
This is equivalent to the fact that the variation of the Hamiltonian was given
simply by a functional variation when $r=0$.

Next, the Lagrangian $(2.2a)$ for the coupling to just a
three-form ($r=1$) with the momentum integrated out becomes
$$
L = { \dot{x}^2 \over 4 } + {1 \over 4}\psi_A\dot{\psi}_A  
  + 3\psi_A\psi_B\dot{x}_C K_{ABC}
  - \psi_A \psi_ B \psi_C \psi_D \partial_A K_{BCD} 
  - 12K_{ABC}K_{ABC} ~ .
\eqno (3.7)
$$
The analogue of (3.5) for the superworldline Lagrangian is
$$
L_s 
\equiv 
\int d\theta \Bigl( {1 \over 4i}DX_AD^2X_A + iDX_ADX_BDX_CK_{ABC} \Bigr) ~ .
\eqno (3.8)
$$
However, although $L_s$ has a simple supersymmetry invariance
$$
\delta_\alpha L_s
\equiv
L_s(x + \delta_\alpha x,\psi + \delta_\alpha \psi) - L_s(x,\psi)
=
\partial_\tau \Bigl[ 
i\alpha\psi_A \bigl( \psi_B\psi_C K_{ABC} 
- {1 \over 4}\dot{x}_A \bigr) \Bigr] ~ ,
\eqno (3.9) 
$$
it does not generate the last term in (3.7), 
$-12K_{ABC}K_{ABC}$
\footnote{${}^\dagger$}{It may be possible to 
generate this term by adding a term to (3.8) where $K_{ABC}$ is contracted 
with a three-form auxiliary superfield, $X_{ABC}$.  However, the auxiliary 
fermionic component of $X_{ABC}$ would have to be constrained so as not to
introduce new degrees of freedom.}.
This means that $L_s$ will not reproduce correct field theory results and
it means that $L$ is not invariant under a simple functional variation 
like that in (3.9).  Nonetheless,
$L$ was shown to be supersymmetric in the previous section when it 
was expressed
in terms of the Hamiltonian.
Notice also that the term $K_{ABC}K_{ABC}$ is exactly the same term 
that similarly 
forbade the Hamiltonian from being supersymmetric simply with  
the Poisson bracket variation.  Finally, as the degree of the form coupled 
to the fermion increases, the superworldline Lagrangian, $L_s$, will fail to 
generate more and more terms that will be present in the 
correct worldline Lagrangian, $L$.

As a consequence of this analysis, consider the Lagrangian for the 
three-form coupling given by (3.7) in dimension $D=6$, which comes
{}from the six dimensional Dirac operator,
$\hat{\Sigma} = \Gamma_A \hat{p}_A - i \Gamma_{ABC} \hat{K}^{(3)}_{ABC}$.  
By defining
the usual axial vector and antisymmetric tensor in four dimensions to 
be $B_\mu \equiv 6K^{(3)}_{\mu 56}$ 
and $K_{\mu\nu} \equiv 3K^{(3)}_{\mu\nu 6}$, respectively, the reduction
to $D=4$ of (3.7) gives
$$
\eqalign{
L
&=
{ \dot{x}^2 \over 4 } + {\bar{x}^2_5 \over 4} + {\bar{x}^2_6 \over 4} 
+ {1\over 4}\psi_\mu\dot{\psi}_\mu 
+ {1\over 4}\psi_5\dot{\psi}_5 + {1\over 4}\psi_6\dot{\psi}_6 
 + \psi_\mu(\bar{x}_5\psi_6 - \bar{x}_6\psi_5)B_\mu
- \bar{x}_6\psi_\mu\psi_\nu K_{\mu\nu}
\cr
&
+ \psi_5\psi_6 \dot{x}_\mu B_\mu
- 2\psi_\mu\psi_6\dot{x}_\nu K_{\mu\nu}
- \psi_\mu\psi_\nu\psi_5\psi_6 \partial_\mu B_\nu 
- \psi_\mu\psi_\nu\psi_\rho\psi_6\partial_\mu K_{\nu\rho} 
- 2B^2 - 4K_{\mu\nu}K_{\mu\nu} ~ ,
\cr}
\eqno (3.10)
$$
where the auxiliary fields are defined as 
$\bar{x}_{5,6} \equiv - \dot{x}_{5,6}$.
The last two terms in (3.10), which are quadratic in the background fields,
are of course the terms which are not predicted by the superworldline 
Lagrangian, $L_s$, of (3.8).  This is why Lagrangian (3.10) is 
presented in the superworldline approach of 
[5] for the axial coupling without the $-2B^2$ term.  It is the 
verdict of this paper that using the worldline Lagrangian (3.10) derived 
{}from field theory, worldline perturbation theory reproduces 
Feynman diagrams, free of any of the ambiguities discussed in [5] on the
treatment of $G_F^2$\footnote{${}^\dagger$}{$G_F(u)$ is the fermion
propagator $\langle \psi (u) \psi (0) \rangle$, which is a step function.}. 
In particular, it is now well understood from [6,7 and 9]
that worldline fermions simply represent $\Gamma$-matrices.  This is the
only fact needed to determine that $G_F^2$ should always be taken to 
be unity {\it even if} it multiplies a $\delta$-function.  The point is 
that by understanding worldline fermions as representing $\Gamma$-matrices,
there is no mixing in the bosonic and fermionic sectors in worldline
perturbation theory.  This assertion has been checked on various
Feynman diagrams.  Similarly, the auxiliary fields may be integrated out
of (3.10), 
but only at the $\Gamma$-level (just like the momentum [6]):
$$
\eqalign{
L
&=
{ \dot{x}^2 \over 4 } 
+ {1\over 4}\psi_\mu\dot{\psi}_\mu 
+ {1\over 4}\psi_5\dot{\psi}_5 + {1\over 4}\psi_6\dot{\psi}_6 
+ \psi_5\psi_6 \dot{x}_\mu B_\mu
- 2\psi_\mu\psi_6\dot{x}_\nu K_{\mu\nu}
- 2K_{\mu\nu}K_{\mu\nu}
\cr
&
- \psi_\mu\psi_\nu\psi_5\psi_6 \partial_\mu B_\nu 
- \psi_\mu\psi_\nu\psi_\rho\psi_6\partial_\mu K_{\nu\rho} 
- 2\psi_\mu\psi_\nu\psi_\rho\psi_5 B_\mu  K_{\nu\rho}
- \psi_\mu\psi_\nu\psi_\rho\psi_\sigma K_{\mu\nu} K_{\rho\sigma} ~ .
\cr}
\eqno (3.11)
$$
This is the exact result obtain in the worldline reformulation of the 
four dimensional Dirac operator
$\hat{\Sigma} 
= \Gamma _\mu \hat{p}_\mu - i\Gamma_\mu\Gamma_5\Gamma_6\hat{B}_\mu 
- i\Gamma _{\mu\nu}\Gamma _6 \hat{K}_{\mu \nu}$.

\bigskip

\noindent
{\bf Acknowledgments}

\bigskip

I gratefully acknowledge helpful conversations with Daniel Cangemi,
Mike M. Cornwall and Eric D'Hoker.

\bigskip

\noindent
{\bf References}

\bigskip

\item{[1]} L. Brink, P. DiVecchia and P. Howe, Nucl. Phys.
           {\bf B118} (1977) 76.
\item{[2]} A.P. Balachandran, P. Salomson, B. Skagerstam and 
           J. Winnberg, Phys. Rev. {\bf D15} (1977) 2308;
           D. Friedan and P. Windey, Nucl. Phys. {\bf B235} (1984) 395.
\item{[3]} E. Witten, Nucl. Phys. {\bf B202} (1982) 253.
\item{[4]} M. Mondrag\'on, L. Nellen, M.G. Schmidt and C. Schubert, 
           Phys. Lett. {\bf 351B} (1995) 200.
\item{[5]} M. Mondrag\'on, L. Nellen, M.G. Schmidt and C. Schubert,  
           Phys. Lett. {\bf B366} (1996) 212.
\item{[6]} E. D'Hoker and D. G. Gagn\'e, Preprint UCLA/95/TEP/35
           (hep-th/9512080), UCLA, 1995 (to appear in Nucl. Phys. {\bf B}).
\item{[7]} E. D'Hoker and D. G. Gagn\'e, Preprint UCLA/95/TEP/22
           (hep-th/9508131), UCLA, 1995 (to appear in Nucl. Phys. {\bf B}).
\item{[8]} R. Casalbuoni, Nuovo Cimento {\bf 33A} (1976) 389.
\item{[9]} M.J. Strassler, Nucl. Phys. {\bf B385} (1992) 145.
 
\end